\documentclass[english]{iopart}
\usepackage{amstext}
\usepackage{graphicx}
\usepackage{amssymb}

\makeatletter

\newcommand{\lyxmathsym}[1]{\ifmmode\begingroup\def\b@ld{bold}
  \text{\ifx\math@version\b@ld\bfseries\fi#1}\endgroup\else#1\fi}

\usepackage{iopams}
\usepackage{setstack}

\usepackage{amsfonts}

\newtheorem{theorem}{Theorem} 
\newtheorem{definition}{Definition} 

\newtheorem{proposition}{Proposition} 

\makeatother

\usepackage{babel}

\begin{document}

\title{Exotic Smoothness and Quantum Gravity II: exotic $\mathbb{R}^{4}$,
singularities and cosmology}

\author{Torsten Asselmeyer-Maluga }

\address{German Aerospace Center, Berlin and }

\address{Loyola University, New Orleans, LA, USA}

\ead{torsten.asselmeyer-maluga@dlr.de}

\author{Jerzy Kr\'ol}

\address{University of Silesia, Institute of Physics, ul. Uniwesytecka 4,
40-007 Katowice, Poland}

\ead{iriking@wp.pl}
\begin{abstract}
Since the first work on exotic smoothness in physics, it was folklore
to assume a direct influence of exotic smoothness to quantum gravity.
In the second paper, we calculate the {}``smoothness structure''
part of the path integral in quantum gravity for the exotic $\mathbb{R}^{4}$
as non-compact manifold. We discuss the influence of the ``sum over
geometries´´ to the ``sum over smoothness structure´´. There are two types of
exotic $\mathbb{R}^{4}$: large (no smooth embedded 3-sphere) and
small (smooth embedded 3-sphere). A large exotic $\mathbb{R}^{4}$
can be produced by using topologically slice but smoothly non-slice
knots whereas a small exotic $\mathbb{R}^{4}$ is constructed by a
5-dimensional h-cobordism between compact 4-manifolds. The results
are applied to the calculation of expectation values, i.e. we discuss
the two observables, volume and Wilson loop. Then the appearance of
naked singularities is analyzed. By using Mostow rigidity, we obtain
a justification of area and volume quantization again. Finally exotic
smoothness of the $\mathbb{R}^{4}$ produces in all cases (small or
large) a cosmological constant.
\end{abstract}


\submitto{\CQG}

\maketitle

\section{Introduction}

Since the first papers about exotic smoothness it was folklore to
state an influence of exotic smoothness on the state sum (or path
integral) for quantum gravity. In our first paper \cite{Ass2010}
we calculated the {}``exotic smoothness'' contribution to the path
integral for a special class of compact 4-manifolds including the
K3 surface. The exotic smoothness structure was constructed by knot
surgery. Similar results were obtained by Duston \cite{Duston2009}
for branched covers. The calculation of the path integral has to formally
include the exotic smoothness \cite{Pfeiffer2004} to relate it to
smooth invariants of 4-manifolds. We demonstrated it in our previous
paper (to get the Chern-Simons invariant). Unfortunately the most
interesting and physically important case of an exotic $\mathbb{R}^{4}$
is also the most complicated one which strongly relies on infinite
constructions (Casson handles etc.). The appearance of two classes
of exotic $\mathbb{R}^{4}$, large and small, complicates the situation.
These two classes have their origin in the two main failures in 4-dimensional
differential topology which stays in contrast to the topological theory:
the smooth h-cobordism theorem and the large class of non-smoothable,
topological 4-manifolds. If there is a smooth embedding of a 3-sphere
into the exotic $\mathbb{R}^{4}$ then one calls it a \emph{small}
exotic $\mathbb{R}^{4}$ and if not it is a \emph{large} exotic $\mathbb{R}^{4}$.
In this paper we will study the effect of the exotic $\mathbb{R}^{4}$
on the functional integral for the Einstein-Hilbert action. 

In the next section we present some of the physical assumptions as
well the definition of an exotic $\mathbb{R}^{4}$. Then we discuss
the existence of a Lorentz metric for an exotic $\mathbb{R}^{4}$.
Formally the existence of a Lorentz metric is a purely topological
question which can be answered positively. On the other hand, global
hyperbolicity and all its consequences depend on the standard smoothness
of $\mathbb{R}^{4}=\mathbb{R}^{3}\times\mathbb{R}$ (as smooth product).
Therefore an exotic $\mathbb{R}^{4}$ must contain naked singularities.
We will further analyze these singularities in section \ref{sec:Naked-singularities}
to obtain a pairwise structure by the failure of the Whitney's trick.
The main part of this paper is formed by the sections \ref{sec:Exotic-R4}
to \ref{sec:The-functional-integral}. It starts with a description
of large and small exotic $\mathbb{R}^{4}$. Then we discuss the splitting
of the action functional (using the diffeomorphism invariance of the
Einstein-Hilbert action) according to these descriptions. Finally
we calculate the functional integral at first for a particular exotic
$\mathbb{R}^{4}$ and then for the whole continuous (radial) family.
The discussion of observables, like volume and the Wilson loop, in
section \ref{sec:Observables} completes the picture. Especially we
confirm the results of Loop quantum gravity, i.e. the quantization
of area and volume. All exotic $\mathbb{R}^{4}$ have one common property:
the appearance of a cosmological constant.

\section{Physical Motivation and model assumptions}

Einsteins insight that gravity is the manifestation of geometry leads
to a new view on the structure of spacetime. From the mathematical
point of view, spacetime is a smooth 4-manifold endowed with a (smooth)
metric as basic variable for general relativity. Later on, the existence
question for Lorentz structure and causality problems (see \cite{HawEll:94})
gave further restrictions on the 4-manifold: causality implies non-compactness,
Lorentz structure needs a codimension-1 foliation. Usually, one starts
with a globally foliated, non-compact 4-manifold $\Sigma\times\mathbb{R}$
fulfilling all restrictions where $\Sigma$ is a smooth 3-manifold
representing the spatial part. But other non-compact 4-manifolds are
also possible, i.e. it is enough to assume a non-compact, smooth 4-manifold
endowed with a codimension-1 foliation. 

All these restrictions on the representation of spacetime by the manifold
concept are clearly motivated by physical questions. Among the properties
there is one distinguished element: the smoothness. Usually one assumes
a smooth, unique atlas of charts covering the manifold where the smoothness
is induced by the unique smooth structure on $\mathbb{R}$. But as
discussed in the introduction, that is not the full story. Even in
dimension 4, there are an infinity of possible other smoothness structures
(i.e. a smooth atlas) non-diffeomorphic to each other. In the following
we will specialize to the $\mathbb{R}^{4}$:

\begin{definition}

The smoothness structure of $\mathbb{R}^{4}$ is called an \textbf{exotic
smoothness structure }or \textbf{exotic $\mathbb{R}^{4}$} if it is
non-diffeomorphic to the standard smoothness structure (induced from
the smooth product $\mathbb{R}\times\mathbb{R}\times\mathbb{R}\times\mathbb{R}$).

\end{definition}

The implications for physics seem to be obvious because we rely on
the smooth calculus to formulate equations of any field theory. Thus
different smoothness structures could represent different physical
situations leading to different measurable results. But it should
be stressed that \emph{exotic smoothness is not exotic physics!} Exotic
smoothness is a mathematical possibility which should be further explored
to understand its physical relevance.

\section{Lorentz metric and global hyperbolicity\label{sec:Lorentz-metric-global-hyp}}

Before we start with the construction of the various exotic $\mathbb{R}^{4}$'s
(large and small), we will discuss some physical implications which
are independent of these constructions. Firstly we consider the existence
of a Lorentz metric, i.e. a 4-manifold $M$ (the spacetime) admits
a Lorentz metric if (and only if) there is a non-vanishing vector
field. In case of a compact 4-manifold $M$ we can use the Poincare-Hopf
theorem to state: a compact 4-manifold admits a Lorentz metric if
the Euler characteristic vanishes $\chi(M)=0$. But in a compact 4-manifold
there are closed time-like curves (CTC) contradicting the causality
or more exactly: the chronology violating set of a compact 4-manifold
is non-empty (Proposition 6.4.2 in \cite{HawEll:94}). Non-compact
4-manifold $M$ admits always a Lorentz metric and a special class
of these 4-manifolds have an empty chronology violating set. If $\mathcal{S}$
is an acausal hypersurface in $M$ (i.e., a topological hypersurface
of $M$ such that no pair of points of $M$ can be connected by means
of a causal curve), then $D^{+}(\mathcal{S})$ is the future Cauchy
development (or domain of dependence) of $\mathcal{S}$, i.e. the
set of all points $p$ of $M$ such that any past-inextensible causal
curve through $p$ intersects $\mathcal{S}$. Similarly $D^{-}(\mathcal{S})$
is the past Cauchy development of $\mathcal{S}$. If there are no
closed causal curves, then $\mathcal{S}$ is a Cauchy surface if $D^{+}(\mathcal{S})\cup\mathcal{S}\cup D^{-}(\mathcal{S})=M$.
But then $M$ is diffeomorphic to $\mathcal{S}\times\mathbb{R}$ \cite{BernalSanchez2003}.
The existence of a Cauchy surface implies global hyperbolicity, i.e.
a spacetime manifold $M$ without boundary is said to be globally
hyperbolic if the following two conditions hold:
\begin{enumerate}
\item \emph{Absence of naked singularities}: For every pair of points $p$
and $q$ in $M$, the space of all points that can be both reached
from $p$ along a past-oriented causal curve and reached from $q$
along a future-oriented causal curve is compact.
\item \emph{Chronology}: No closed causal curves exist (or ''Causality''
holds on $M$).
\end{enumerate}
Usually condition 2 above is replaced by the more technical condition
''Strong causality holds on $M$'' but as
shown in \cite{BernalSanchez2007} instead of ''strong
causality'', one can write simply the condition ''causality''
(and strong causality will hold under causality plus condition 1 above). 

Then all (non-compact) 4-manifolds $\mathcal{S}\times\mathbb{R}$
are the only 4-manifolds which admit a globally hyperbolic Lorentz
metric, where the product $\times$ has to be a smooth product not
only by physical reasons but also because of the claimed result in
\cite{BernalSanchez2003}. But more is true \cite{BernalSanchez2005}, 

\begin{theorem}\label{thm:global-hyp}

If a spacetime $(M,g)$ is globally hyperbolic, then it is isometric
to $(\mathbb{R}\times\mathcal{S},-f\cdot d\tau^{2}+g_{\tau})$
with a smooth positive function $f:\mathbb{R}\to\mathbb{R}$ and a
smooth family of Riemannian metrics $g_{\tau}$ on $\mathcal{S}$
varying with $\tau$. Moreover, each $\left\{ t\right\} \times\mathcal{S}$
is a Cauchy slice.

\end{theorem} Furthermore in \cite{BernalSanchez2006} it was shown:
\begin{itemize}
\item If a compact spacelike submanifold with boundary of a globally hyperbolic
spacetime is acausal then it can be extended to a full Cauchy spacelike
hypersurface $\mathcal{S}$ of $M$, and
\item for any Cauchy spacelike hypersurface $\mathcal{S}$ there exists
a function as in Th. \ref{thm:global-hyp} such that $\mathcal{S}$
is one of the levels $\tau=constant$.
\end{itemize}
So, what about exotic 4-manifolds? At first the existence of the Lorentz
metric is a purely topological condition which will be fulfilled by
all non-compact 4-manifolds independent of the smoothness structure.
By considering the global hyperbolicity, the picture changes. An exotic
spacetime homeomorphic to $\mathcal{S}\times\mathbb{R}$ is not diffeomorphic
to $\mathcal{S}\times\mathbb{R}$. The Cauchy surface $\mathcal{S}$
is a 3-manifold with unique smoothness structure (up to diffeomorphisms),
the standard structure. So, the smooth product $\mathcal{S}\times\mathbb{R}$
has also the standard smoothness structure. But the diffeomorphism
to $\mathcal{S}\times\mathbb{R}$ is necessary for global hyperbolicity.
Therefore an \emph{exotic $\mathcal{S}\times\mathbb{R}$ is never
globally hyperbolic but admits a Lorentz metric}. Generally we have
an exotic $\mathcal{S}\times\mathbb{R}$ with a Lorentz metric such
that the projection $\mathcal{S}\times\mathbb{R}\to\mathbb{R}$ is
a time-function (that is, a ontinuous function which is strictly increasing
on future directed causal curves). But then the exotic $\mathcal{S}\times\mathbb{R}$
has no closed causal curves and must contain naked singularities%
\footnote{Any non-compact manifold $M$ admits stably causal metrics (that is,
those with a time function). So, if $M$ is not diffeomorphic to some
product$\mathcal{S}\times\mathbb{R}$, all these (causally well behaved)
metrics must contain naked singularities. We thank M. S\'anchez for
the explanation of this result.%
}. We will later see the source of these singularities.

\section{Exotic $\mathbb{R}^{4}$\label{sec:Exotic-R4}}

In this section we will give some information about the construction
of exotic $\mathbb{R}^{4}$. The existence of a smooth embedding $S^{3}\to R^{4}$
in the exotic $\mathbb{R}^{4}$ splits all exotic $\mathbb{R}^{4}$
into two classes, large (no embedding) or small.

\subsection{Preliminaries: Slice and non-slice knots}

\begin{figure}
\includegraphics[scale=0.5]{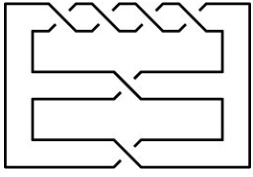}

\caption{Stevedore's knot $6_{1}$\label{fig:Stevedore-knot-6_1}}

\end{figure}

\begin{figure}
\includegraphics[angle=90]{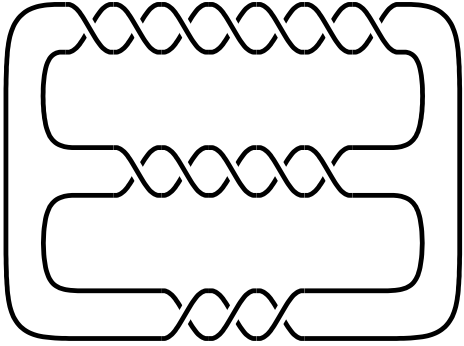}

\caption{pretzel knot $(-3,5,7)$\label{fig:pretzel-knot-3-5-7}}

\end{figure}

At first we start with some definitions from knot theory. A (smooth)
knot $K$ is a smooth embedding $S^{1}\to S^{3}$. In the following
we assume every knot to be smooth. Secondly we exclude wilderness
of knots, i.e the knot is equivalent to a polygon in $\mathbb{R}^{3}$
or $S^{3}$ (tame knot). Furthermore, the $n$-disk is denoted by
$D^{n}$ with $\partial D^{n}=S^{n-1}$. \begin{definition} \textbf{Smoothly
Slice Knot}: A knot in $\partial D^{4}=S^{3}$ is smoothly slice if
there exists a two-disk $D^{2}$ smoothly embedded in $D^{4}$ such
that the image of $\partial D^{2}=S^{1}$ is $K$. \end{definition}
An example of a slice knot is the so-called Stevedore's Knot (in Rolfson
notation $6_{1}$, see Fig. \ref{fig:Stevedore-knot-6_1}).%
 \begin{definition} \textbf{Flat Topological Embedding}: Let $X$
be a topological manifold of dimension $n$ and $Y$ a topological
manifold of dimension $m$ where $n<m$. A topological embedding $\rho:X\to Y$
is flat if it extends to a topological embedding $\rho:X\times D^{m-n}\to Y$. 

\textbf{Topologically Slice Knot}: A knot $K$ in $\partial D^{4}$
is topologically slice if there exists a two-disk $D^{2}$ flatly
topologically embedded in $D^{4}$ such that the image of $\partial D^{2}$
is $K$. \end{definition} Here we remark that the flatness condition
is essential. Any knot $K\subset S^{3}$ is the boundary of a disc
$D^{2}$ embedded in $D^{4}$, which can be seen by taking the cone
over the knot. But the vertex of the cone is a non-flat point (the
knot is crashed to a point). The difference between the smooth and
the flat topological embedding is the key for the following discussion.
This innocent looking difference seem to imply that both definitions
are equivalent. But deep results from 4-manifold topology gave a negative
answer: there are topologically slice knots which are not smoothly
slice. An example is the pretzel knot $(-3,5,7)$ (see Fig. \ref{fig:pretzel-knot-3-5-7}).
 In \cite{Fre:82a}, Freedman gave a topological criteria for topological
sliceness: the Alexander polynomial $\triangle_{K}(t)$ (the best
known knot invariant, see \cite{Rol:76}) of the knot $K$ has to
be one, $\triangle_{K}(t)=1$. An example how to measure the smooth
sliceness is given by the smooth 4-genus $g_{4}(K)$ of the knot $K$,
i.e. the minimal genus of a surface $F$ smoothly embedded in $D^{4}$
with boundary $\partial F=K$ the knot. This surface $F$ is called
the \emph{Seifert surface}. Therefore, if the smooth 4-genus vanishes
$g_{4}(K)=0$ then the knot $K$ bounds a 2-disk $D^{2}$ (surface
of genus $0$) given by the smooth embedding $D^{2}\to D^{4}$ so
that the image of $\partial D^{2}\to\partial D^{4}$ is the knot $K$.

\subsection{Large exotic $\mathbb{R}^{4}$ and non-slice knots\label{sub:Large-exotic-R4}}

Large exotic $\mathbb{R}^{4}$ can be constructed by using the failure
to arbitrarily split of a compact, simple-connected 4-manifold. For
every topological 4-manifold one knows how to split this manifold
\emph{topologically} into simpler pieces using the work of Freedman
\cite{Fre:82}. But as shown by Donaldson \cite{Don:83}, some of
these 4-manifolds do not exist as smooth 4-manifolds. This contradiction
between the continuous and the smooth case produces the first examples
of exotic $\mathbb{R}^{4}$\cite{Gom:83}. Unfortunately, the construction
method is rather indirect and therefore useless for the calculation
of the path integral contribution of the exotic $\mathbb{R}^{4}$.
But as pointed out by Gompf (see \cite{Gom:85} or \cite{GomSti:97}
Exercise 9.4.23 on p. 377ff and its solution on p. 522ff), large exotic
$\mathbb{R}^{4}$ can be also constructed by using smoothly non-slice
but topologically slice knots. Especially one obtains an explicit
construction which will be used in the calculations later.

Let $K$ be a knot in $\partial D^{4}$ and $X_{K}$ the two-handlebody
obtained by attaching a two-handle to $D^{4}$ along $K$ with framing
$0$. That means: one has a two-handle $D^{2}\times D^{2}$ which
is glued to the 0-handle $D^{4}$ along its boundary using a map $f:\partial D^{2}\times D^{2}\to\partial D^{4}$
so that $f(.\,,\, x)=K\times x\subset S^{3}=\partial D^{4}$ for all
$x\in D^{2}$ (or the image $im(f)=K\times D^{2}$ is the solid knotted
torus). Let $\rho:X_{K}\to\mathbb{R}^{4}$ be a flat topological embedding
($K$ is topologically slice). For $K$ a smoothly non-slice knot,
the open 4-manifold\begin{equation}
R^{4}=\left(\mathbb{R}^{4}\setminus int\rho(X_{K})\right)\cup_{\partial X_{K}}X_{K}\label{eq:decomposition-large-exotic-R4}\end{equation}
where $int\rho(X_{K})$ is the interior of $\rho(X_{K})$, is homeomorphic
but non-diffeomorphic to $\mathbb{R}^{4}$ with the standard smoothness
structure (both pieces are glued along the common boundary $\partial X_{K}$).
The proof of this fact ($R^{4}$ is exotic) is given by contradiction,
i.e. let us assume $R^{4}$ is diffeomorphic to $\mathbb{R}^{4}$.
Thus, there exists a diffeomorphism $R^{4}\to\mathbb{R}^{4}$. The
restriction of this diffeomorphism to $X_{K}$ is a smooth embedding
$X_{K}\hookrightarrow\mathbb{R}^{4}$. However, such a smooth embedding
exists if and only if $K$ is smoothly slice (see \cite{GomSti:97}).
But, by hypothesis, $K$ is not smoothly slice. Thus by contradiction,
there exists a no diffeomorphism $R^{4}\to\mathbb{R}^{4}$ and $R^{4}$
is exotic, homeomorphic but not diffeomorphic to $\mathbb{R}^{4}$.
Finally, we have to prove that $R^{4}$ is large. $X_{K}$, by construction,
is compact and a smooth submanifold of $R^{4}$. By hypothesis, $K$
is not smoothly slice and therefore $X_{K}$ can not smoothly embed
in $\mathbb{R}^{4}$. By restriction, $D^{4}\subset X_{K}$ and also
$\partial D^{4}=S^{3}$ can not smoothly embed and therefore $R^{4}$
is a large exotic $\mathbb{R}^{4}$.

\subsection{Small exotic $\mathbb{R}^{4}$ and Casson handles\label{sub:Small-exotic-R4}}

Small exotic $\mathbb{R}^{4}$'s are again the result of anomalous
smoothness in 4-dimensional topology but of a different kind than
for large exotic $\mathbb{R}^{4}$'s. In 4-manifold topology \cite{Fre:82},
a homotopy-equivalence between two compact, closed, simply-connected
4-manifolds implies a homeomorphism between them (a so-called h cobordism).
But Donaldson \cite{Don:87} provided the first smooth counterexample,
i.e. both manifolds are generally not diffeomorphic to each other.
The failure can be localized in some contractible submanifold (Akbulut
cork) so that an open neighborhood of this submanifold is a small
exotic $\mathbb{R}^{4}$. The whole procedure implies that this exotic
$\mathbb{R}^{4}$ can be embedded in the 4-sphere $S^{4}$.

The idea of the construction is simply given by the fact that every
such smooth h-cobordism between non-diffeomorphic 4-manifolds can
be written as a product cobordism except for a compact contractible
sub-h-cobordism $V$, the Akbulut cork. An open subset $U\subset V$
homeomorphic to $[0,1]\times{{\mathbb{R}}^{4}}$ is the corresponding
sub-h-cobordism between two exotic ${{\mathbb{R}}^{4}}$'s. These
exotic ${{\mathbb{R}}^{4}}$'s are called ribbon ${{\mathbb{R}}^{4}}$'s.
They have the important property of being diffeomorphic to open subsets
of the standard ${{\mathbb{R}}^{4}}$. To be more precise, consider
a pair $(X_{+},X_{-})$ of homeomorphic, smooth, closed, simply-connected
4-manifolds. 

\begin{theorem}\emph{ }Let $W$ be a smooth h-cobordism between closed,
simply connected 4-manifolds $X_{-}$ and $X_{+}$. Then there is
an open subset $U\subset W$ homeomorphic to $[0,1]\times{{\mathbb{R}}^{4}}$
with a compact subset $C\subset U$ such that the pair $(W\setminus C,U\setminus C)$
is diffeomorphic to a product $[0,1]\times(X_{-}\setminus C,U\cap X_{-}\setminus C)$.
The subsets $R_{\pm}=U\cap X_{\pm}$ (homeomorphic to ${{\mathbb{R}}^{4}}$)
are diffeomorphic to open subsets of ${{\mathbb{R}}^{4}}$. If $X_{-}$
and $X_{+}$ are not diffeomorphic, then there is no smooth 4-ball
in $R_{\pm}$ containing the compact set $Y_{\pm}=C\cap R_{\pm}$,
so both $R_{\pm}$ are exotic ${{\mathbb{R}}^{4}}$'s.\emph{ }\end{theorem}

Thus, remove a certain contractible, smooth, compact 4-manifold $Y_{-}\subset X_{-}$
(called an Akbulut cork) from $X_{-}$, and re-glue it by an involution
of $\partial Y_{-}$, i.e. a diffeomorphism $\tau:\partial Y_{-}\to\partial Y_{-}$
with $\tau\circ\tau=Id$ and $\tau(p)\not=\pm p$ for all $p\in\partial Y_{-}$.
This argument was modified above so that it works for a contractible
{\em open} subset $R_{-}\subset X_{-}$ with similar properties,
such that $R_{-}$ will be an exotic ${{\mathbb{R}}^{4}}$ if $X_{+}$
is not diffeomorphic to $X_{-}$. Furthermore $R_{-}$ lies in a compact
set, i.e. a 4-sphere or $R_{-}$ is a small exotic $\mathbb{R}^{4}$.
In the next subsection we will see how this results in the construction
of handlebodies of exotic ${{\mathbb{R}}^{4}}$. In \cite{DeMichFreedman1992}
Freedman and DeMichelis constructed also a continuous family of small
exotic $\mathbb{R}^{4}$.

Now we are ready to discuss the decomposition of a small exotic $\mathbb{R}^{4}$
by Bizaca and Gompf \cite{BizGom:96} by using special pieces, the
handles forming a handle body. Every 4-manifold can be decomposed
(seen as handle body) using standard pieces such as $D^{k}\times D^{4-k}$,
the so-called $k$-handle attached along $\partial D^{k}\times D^{4-k}$
to the boundary $S^{3}=\partial D^{4}$ of a $0-$handle $D^{0}\times D^{4}=D^{4}$.
The construction of the handle body can be divided into two parts.
The first part is known as the Akbulut cork, a contractable 4-manifold
with boundary a homology 3-sphere (a 3-manifold with the same homology
as the 3-sphere). The Akbulut cork $A_{cork}$ is given by a linking
between a 1-handle and a 2-handle of framing $0$. The second part
is the Casson handle $CH$ which will be considered now.

Let us start with the basic construction of the Casson handle $CH$.
Let $M$ be a smooth, compact, simple-connected 4-manifold and $f:D^{2}\to M$
a (codimension-2) mapping. By using diffeomorphisms of $D^{2}$ and
$M$, one can deform the mapping $f$ to get an immersion (i.e. injective
differential) generically with only double points (i.e. $\#|f^{-1}(f(x))|=2$)
as singularities \cite{GolGui:73}. But to incorporate the generic
location of the disk, one is rather interesting in the mapping of
a 2-handle $D^{2}\times D^{2}$ induced by $f\times id:D^{2}\times D^{2}\to M$
from $f$. Then every double point (or self-intersection) of $f(D^{2})$
leads to self-plumbings of the 2-handle $D^{2}\times D^{2}$. A self-plumbing
is an identification of $D_{0}^{2}\times D^{2}$ with $D_{1}^{2}\times D^{2}$
where $D_{0}^{2},D_{1}^{2}\subset D^{2}$ are disjoint sub-disks of
the first factor disk%
\footnote{In complex coordinates the plumbing may be written as $(z,w)\mapsto(w,z)$
or $(z,w)\mapsto(\bar{w},\bar{z})$ creating either a positive or
negative (respectively) double point on the disk $D^{2}\times0$ (the
core).%
}. Consider the pair $(D^{2}\times D^{2},\partial D^{2}\times D^{2})$
and produce finitely many self-plumbings away from the attaching region
$\partial D^{2}\times D^{2}$ to get a kinky handle $(k,\partial^{-}k)$
where $\partial^{-}k$ denotes the attaching region of the kinky handle.
A kinky handle $(k,\partial^{-}k)$ is a one-stage tower $(T_{1},\partial^{-}T_{1})$
and an $(n+1)$-stage tower $(T_{n+1},\partial^{-}T_{n+1})$ is an
$n$-stage tower union kinky handles $\bigcup_{\ell=1}^{n}(T_{\ell},\partial^{-}T_{\ell})$
where two towers are attached along $\partial^{-}T_{\ell}$. Let $T_{n}^{-}$
be $(\mbox{interior}T_{n})\cup\partial^{-}T_{n}$ and the Casson handle
\[
CH=\bigcup_{\ell=0}T_{\ell}^{-}\]
is the union of towers (with direct limit topology induced from the
inclusions $T_{n}\hookrightarrow T_{n+1}$). 

The main idea of the construction above is very simple: an immersed
disk (disk with self-intersections) can be deformed into an embedded
disk (disk without self-intersections) by sliding one part of the
disk along another (embedded) disk to kill the self-intersections.
Unfortunately the other disk can be immersed only. But the immersion
can be deformed to an embedding by a disk again etc. In the limit
of this process one ''shifts the self-intersections into infinity''
and obtains%
\footnote{In the proof of Freedman \cite{Fre:82}, the main complications come
from the lack of control about this process. %
} the standard open 2-handle $(D^{2}\times\mathbb{R}^{2},\partial D^{2}\times\mathbb{R}^{2})$. 

A Casson handle is specified up to (orientation preserving) diffeomorphism
(of pairs) by a labeled finitely-branching tree with base-point {*},
having all edge paths infinitely extendable away from {*}. Each edge
should be given a label $+$ or $-$. Here is the construction: tree
$\to CH$. Each vertex corresponds to a kinky handle; the self-plumbing
number of that kinky handle equals the number of branches leaving
the vertex. The sign on each branch corresponds to the sign of the
associated self plumbing. The whole process generates a tree with
infinite many levels. In principle, every tree with a finite number
of branches per level realizes a corresponding Casson handle. Each
building block of a Casson handle, the {}``kinky'' handle with $n$
kinks%
\footnote{The number of end-connected sums is exactly the number of self intersections
of the immersed two handle.%
}, is diffeomorphic to the $n-$times boundary-connected sum $\natural_{n}(S^{1}\times D^{3})$
(see appendix \ref{sec:Connected-and-boundary-connected}) with two
attaching regions. Technically speaking, one region is a tubular neighborhood
of band sums of Whitehead links connected with the previous block.
The other region is a disjoint union of the standard open subsets
$S^{1}\times D^{2}$ in $\#_{n}S^{1}\times S^{2}=\partial(\natural_{n}S^{1}\times D^{3})$
(this is connected with the next block).

\section{The action functional\label{sec:action-functional}}

In this section we will discuss the Einstein-Hilbert action functional
\begin{equation}
S_{EH}(M)=\intop_{M}R\sqrt{g}\: d^{4}x\label{eq:EH-action}\end{equation}
of the 4-manifold $M$ and fix the Ricci-flat metric $g$ as solution
of the vacuum field equations of the exotic 4-manifold. The main part
of our argumentation is additional contribution to the action functional
coming from exotic smoothness.

\subsection{Large exotic $\mathbb{R}^{4}$}

In case of the large exotic $\mathbb{R}^{4}$, we consider the decompositions\begin{eqnarray}
R^{4} & = & \left(\mathbb{R}^{4}\setminus int\rho(X_{K})\right)\cup_{\partial X_{K}}X_{K}\label{eq:relation-exotic}\\
\mathbb{R}^{4} & = & \left(\mathbb{R}^{4}\setminus int\rho(X_{K})\right)\cup_{\partial X_{K}}\rho(X_{K})\label{eq:relation-standard}\end{eqnarray}
leading to a sum in the action \begin{eqnarray*}
S_{EH}(\mathbb{R}^{4})=\intop_{\mathbb{R}^{4}}R\sqrt{g}\: d^{4}x & = & \intop_{\mathbb{R}^{4}\setminus int\rho(X_{K})}R\sqrt{g}\: d^{4}x+\intop_{\rho(X_{K})}R\sqrt{g}\: d^{4}x\\
 & = & S_{EH}(\mathbb{R}^{4}\setminus int\rho(X_{K}))+S_{EH}(\rho(X_{K}))\quad.\end{eqnarray*}
Because of diffeomorphism invariance of the Einstein-Hilbert action,
this decomposition do not depend on the concrete realization with
respect to any coordinate system. Therefore we obtain the relation\begin{equation}
S_{EH}(\mathbb{R}^{4}\setminus int\rho(X_{K}))=S_{EH}(\mathbb{R}^{4})-S_{EH}(\rho(X_{K}))\label{eq:relation-1}\end{equation}
and get a similar relation using (\ref{eq:relation-exotic}) between
the action $S_{EH}(\mathbb{R}^{4})$ of the standard $\mathbb{R}^{4}$
and the action $S_{EH}(R^{4})$ of the large exotic $\mathbb{R}^{4}$\begin{eqnarray}
S_{EH}(R^{4}) & = & S_{EH}(\mathbb{R}^{4}\setminus int\rho(X_{K}))+S_{EH}(X_{K})\nonumber \\
 & = & S_{EH}(\mathbb{R}^{4})+S_{EH}(X_{K})-S_{EH}(\rho(X_{K}))\quad.\label{eq:action-relation-1}\end{eqnarray}
The knot is topologically slice ($\rho$ is a flat topological embedding).
Therefore the restriction of $\rho$ to the 2-handle $D^{2}\times D^{2}$
in $X_{K}$ is a topological embedding defining an embedding $\rho':D^{2}\to D^{4}$
with $\rho'(\partial D^{2})=K$. From the topological point of view,
the Seifert surface of $K$ is the disc $D^{2}$ with genus $0$.
Then we obtain using $X_{K}=D^{4}\cup(D^{2}\times D^{2})$\[
S_{EH}(\rho(X_{K}))=S_{EH}(\rho(D^{4}))+S_{EH}(\rho(D^{2}\times D^{2}))=S_{EH}(\rho(D^{2}\times D^{2}))\]
assuming the flatness of the $0-$handle $D^{4}$. The product metric
(block diagonal metric) \[
ds^{2}=g_{D_{1}}dx_{1}^{2}+g_{D_{2}}dx_{2}^{2}\]
on the image $\rho(D^{2}\times D^{2})=D_{1}\times D_{2}$ of the 2-handle
with $\partial D_{1}=K$ induces\[
S_{EH}(\rho(D^{2}\times D^{2}))=vol(D_{2})\cdot\intop_{D_{1}}R_{D_{1}}\sqrt{g_{D_{1}}}d^{2}x_{1}+vol(D_{1})\cdot\intop_{D_{2}}R_{D_{2}}\sqrt{g_{D_{2}}}d^{2}x_{2}\]
with the curvature scalars $R_{D_{1}},R_{D_{2}}$. The 2-dimensional
integrals\begin{eqnarray*}
\intop_{D_{1}}R_{D_{1}}\sqrt{g_{D_{1}}}d^{2}x_{1} & = & 2\pi\cdot\chi(D_{1})=2\pi\\
\intop_{D_{2}}R_{D_{2}}\sqrt{g_{D_{2}}}d^{2}x_{2} & = & 2\pi\cdot\chi(D_{2})=2\pi\end{eqnarray*}
are by definition the Euler characteristics $\chi(D_{1})=1,\chi(D_{2})=1$
using the topologically sliceness of the knot $K=\partial D_{1}$.
Finally we obtain\[
S_{EH}(\rho(X_{K}))=2\pi\cdot vol(D_{2})+2\pi\cdot vol(D_{1})\quad.\]
Now we consider the other action $S_{EH}(X_{K})$ where we use a non-flat
embedding $X_{K}\hookrightarrow\mathbb{R}^{4}$. Remember the knot
$K$ is smoothly not slice. But then we can only choose the embedding
so that the minimal genus $g_{4}(K)$ of the Seifert surface $F$
is non-zero, i.e. one obtains for the Euler characteristics\[
\intop_{F}R_{F}\sqrt{g_{F}}d^{2}x=2\pi\cdot(1-2g_{4}(K))\quad.\]
This genus $g_{4}(K)$ is an invariant of the knot also known as smooth
4-genus. Importantly the Seifert surface $F$ has negative curvature
for $g_{4}(K)>0$. A similar argumentation leads to the result\[
S_{EH}(X_{K})=2\pi\cdot vol(D_{2})\cdot(1-2g_{4}(K))+2\pi\cdot vol(F)\]
and finally we have the relation using (\ref{eq:action-relation-1})
and the results above\begin{equation}
S_{EH}(R^{4})=S_{EH}(\mathbb{R}^{4})-4\pi\cdot vol(D_{2})\cdot g_{4}(K)+2\pi\cdot(vol(F)-vol(D_{1}))\label{eq:relation-actions-2}\end{equation}
as the correction to the action $S_{EH}(R^{4})$ of the large exotic
$\mathbb{R}^{4}$. The two surfaces $F$ and $D_{1}$ have the same
boundary (the knot $K$) and differ only by the embedding. So, it
seems natural to assume the same volume, i.e. $vol(F)=vol(D_{1})$.
Finally we will write this relation in the usual units\begin{equation}
\frac{1}{\hbar}S_{EH}(R^{4})=\frac{1}{\hbar}S_{EH}(\mathbb{R}^{4})-\frac{vol(D_{2})}{L_{P}^{2}}\cdot4\pi^{2}\cdot g_{4}(K)\label{eq:relation-actions-large-exotic-R4}\end{equation}
This expression looks very simple but the complication is located
at the 4-genus $g_{4}(K)$. Currently there is no simple expression
for the calculation. All results show only the existence $g_{4}(K)\not=0$
but never calculate the value. So, we are not satisfied with the expression
above. We would expect that $g_{4}(K)$ is related to the map $\rho$
which is certainly related to infinite constructions like the Casson
handle. If this speculation is correct then one can interpret the
expression above as a non-perturbative calculation.

\subsection{Small exotic $\mathbb{R}^{4}$}

\begin{figure}
\includegraphics[scale=0.2]{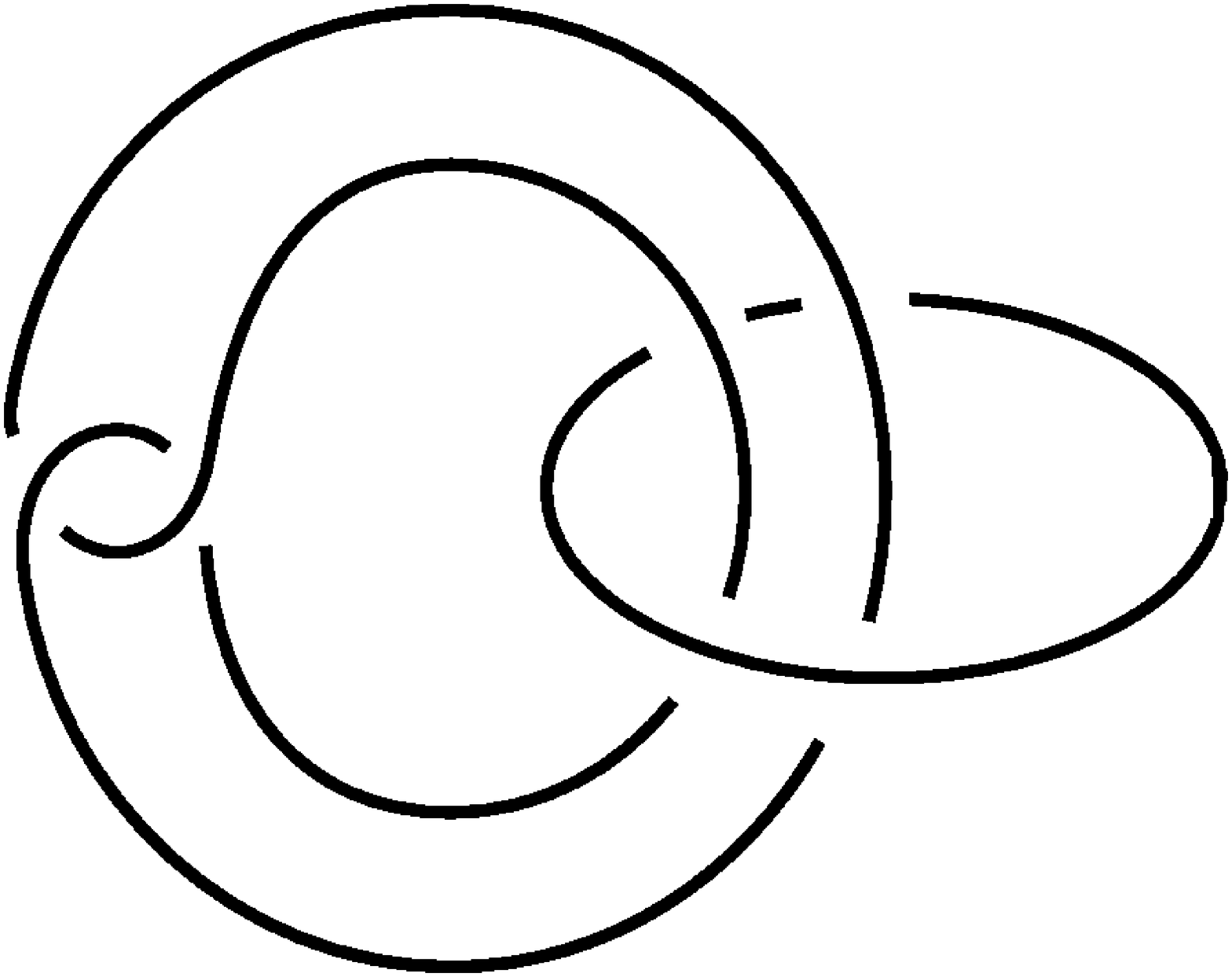}

\caption{Whitehead link $Wh$ \label{fig:Whitehead-link}}

\end{figure}

As explained above, a small exotic $\mathbb{R}^{4}$ can be decomposed
into a compact subset $A_{cork}$ (Akbulut cork) and a Casson handle
(see \cite{BizGom:96}). Especially this exotic $\mathbb{R}^{4}$
depends strongly on the Casson handle, i.e. non-diffeomorphic Casson
handles lead to non-diffeomorphic $\mathbb{R}^{4}$'s. Thus we have
to understand the analytical properties of a Casson handle. In \cite{Kato2004},
the analytical properties of the Casson handle were discussed. The
main idea is the usage of the theory of end-periodic manifolds, i.e.
an infinite periodic structure generated by $W$ glued along a compact
set $A_{cork}$ to get for the interior \[
\mathbb{R}_{\theta}^{4}=int\left(A_{cork}\cup_{N}W\cup_{N}W\cup_{N}\cdots\right)\]
the end-periodic manifold. The definition of an end-periodic manifold
is very formal (see \cite{Tau:87}) and we omit it here. All Casson
handles generated by a balanced tree have the structure of end-periodic
manifolds as shown in \cite{Kato2004}. By using the theory of Taubes
\cite{Tau:87} one can construct a metric on $\cdots\cup_{N}W\cup_{N}W\cup_{N}\cdots$
by using the metric on $W$. Then a metric $g$ in $\mathbb{R}_{\theta}^{4}$
transforms to a periodic function $\hat{g}$ on the infinite periodic
manifold\[
\tilde{Y}=\cdots\cup_{N}W_{-1}\cup_{N}W_{0}\cup_{N}W_{1}\cup_{N}\cdots\]
 where $W_{i}$ is the building block $W$ at the $i$th place. Then
the action of $\mathbb{R}_{\theta}^{4}$ can be divided into two parts\begin{equation}
S_{EH}(\mathbb{R}_{\theta}^{4})=S_{EH}(A_{cork})+\sum_{i}S_{EH}(W_{i})\label{eq:action-exotic-R4-1}\end{equation}
and we start with the discussion of the compact part $A_{cork}$.
This part $A_{cork}$ is formally given by a so-called plumbing of
two spheres $A,B$ with trivial normal bundles having the algebraic
intersection number $A\cdot B=1$ but an extra pair of intersections
(with numbers $+1$ and $-1$). The whole construction can be simplified
(see \cite{GomSti:97} p. 361ff) to obtain a diffeomorphism to the
Akbulut cork. The boundary of the cork is a homology 3-sphere (a Brieskorn
sphere $\Sigma(2,5,7)$ see \cite{AkbKir:79}) with metric of constant
curvature. Without loss of generality we choose a homogeneous metric
in the interior of the cork $int(A_{cork})$ as well and obtain a
constant action\begin{equation}
S_{EH}(int(A_{cork}))=\lambda_{A_{cork}}\cdot vol(A_{cork})\label{eq:action-K}\end{equation}
with respect to the volume of $A_{cork}$ and the curvature $\lambda_{A_{cork}}$.
Because of the non-trivial attaching region, the action for $W_{i}$
has to be non-trivial too. As explained above the simplest part of
a Casson handle is the kinky handle given by the $n-$times boundary-connected
sum $\natural_{n}(S^{1}\times D^{3})$ where $n$ is the number of
self-intersections. Here we will discuss the simplest case $Tree_{+}$
of a Casson handle with one self-intersection at each level first.
It is known by the work of Bizaca \cite{Biz:95} that this Casson
handle admits an exotic smoothness structure. Therefore we assume
now a kinky handle with one self-intersection. The attaching region
is the disjoint union of $S^{1}\times D^{2}$ which are glued together
along $S^{1}\times\partial D^{2}$ to form the boundary $S^{1}\times S^{2}=\partial(S^{1}\times D^{3})$.
Then the attaching to the boundary of the Akbulut cork $\partial A_{cork}$
is given by a map $\phi:\left(S^{1}\times D^{2}\right)\sqcup\left(S^{1}\times D^{2}\right)\to\partial A_{cork}$
where $\left(S^{1}\times D^{2}\right)\cup\left(S^{1}\times D^{2}\right)$
is mapped to a thickened Whitehead link $N(Wh)$ with \begin{eqnarray*}
N(Wh) & = & Wh\times D^{2}=\phi\left(\left(S^{1}\times D^{2}\right)\cup\left(S^{1}\times D^{2}\right)\right)\\
 & = & \phi(S^{1}\times S^{2})=\phi(\partial(S^{1}\times D^{3}))\subset\partial A_{cork}\end{eqnarray*}
(see Fig. \ref{fig:Whitehead-link} for $Wh$). %
 Technically speaking, one attaches two 2-handles along the Whitehead
link. Usually one has to define the framing of the link (as degree
of $\phi$). The value of the action for $S^{1}\times D^{3}$ is unimportant
because the main information is contained in the attaching map $\phi$
and the action $S_{EH}(W_{1})$ (where $W_{1}$ is attached to $A_{cork}$)
is given by\[
S_{EH}(W_{1})=\intop_{\phi(\partial(S^{1}\times D^{3}))\times(0,\epsilon)}R\sqrt{g}d^{4}x\]
with an epsilon neighborhood of the attaching map $\phi(\partial(S^{1}\times D^{3}))\times(0,\epsilon)$.
This neighborhood is necessary to represent the framing of the Whitehead
link. In our case this framing is zero and we obtain an epsilon neighborhood
as product. The introduction of a product metric for this neighborhood\[
ds^{2}=d\theta^{2}+h_{ik}dx^{i}dx^{k}\]
with coordinate $\theta$ on $(0,\epsilon)$ and metric $h_{ik}$
on the attaching region $\phi(\partial(S^{1}\times D^{3}))\subset\partial A_{cork}$.
We are using the ADM formalism with the lapse $N$ and shift function
$N^{i}$ to get a relation between the 4-dimensional $R$ and the
3-dimensional scalar curvature $R_{(3)}$ (see \cite{MiThWh:73} (21.86)
p. 520)\[
\sqrt{g}\, R\: d^{4}x=N\sqrt{h}\:\left(R_{(3)}+||n||^{2}((tr\mathbf{K})^{2}-tr\mathbf{K}^{2})\right)d\theta\, d^{3}x\]
with the normal vector $n$ and the extrinsic curvature $\mathbf{K}$.
We can arrange that the extrinsic curvature has a fixed value $\mathbf{K}=const.$
Then we obtain\[
\intop_{\phi(\partial(S^{1}\times D^{3}))\times(0,\epsilon)}R\sqrt{g}d^{4}x=\epsilon\intop_{\phi(\partial(S^{1}\times D^{3}))}R_{(3)}N\sqrt{h}d^{3}x\]
with the integral $\epsilon=\int d\theta$. As mentioned above, because
of the framing the epsilon do not vanish and we choose a minimal length
$\epsilon=L_{P}$. Finally we have\[
S_{EH}(W_{1})=L_{P}\intop_{\phi(\partial(S^{1}\times D^{3}))}R_{(3)}N\sqrt{h}d^{3}x=L_{P}\intop_{N(Wh)}R_{(3)}N\sqrt{h}d^{3}x\]
and we integrate over the thickened Whitehead link $N(Wh)\subset\partial A_{cork}$.
Now we use the standard trick\[
\partial A_{cork}=(\partial A_{cork}\setminus N(Wh))\cup N(Wh)\]
to express $N(Wh)$ by $\partial A_{cork}$ and the link complement
$\partial A_{cork}\setminus N(Wh)$. Then we obtain for the action\[
S_{EH}(W_{1})=L_{P}\left(\intop_{\partial A_{cork}}R_{(3)}N\sqrt{h}d^{3}x-\intop_{\partial A_{cork}\setminus N(Wh)}R_{(3)}N\sqrt{h}d^{3}x\right)\]
and we have to deal with 3-dimensional Einstein-Hilbert action over
$\partial A_{cork}$ and the link complement $\partial A_{cork}\setminus N(Wh)$
only. But as shown by Witten \cite{Wit:89.2,Wit:89.3,Wit:91.2} this
integral \[
\intop_{\Sigma}R_{(3)}N\sqrt{h}d^{3}x=L_{\Sigma}\cdot CS(\Sigma)\]
over the 3-manifold $\Sigma$ is the Chern-Simons invariant of $\Sigma$.
The length is set to $L_{\Sigma}=\sqrt[3]{vol(\Sigma)}$. Thus the
action is given by\begin{equation}
S_{EH}(W_{1})=L_{P}\left(\sqrt[3]{vol(\partial A_{cork})}\cdot CS(\partial A_{cork})-\sqrt[3]{vol(\partial A_{cork}\setminus N(Wh))}\cdot CS(\partial A_{cork}\setminus N(Wh))\right)\label{eq:action-W1}\end{equation}
for the first level. Beginning with the next level, we have the attachment
of the kinky handle to the boundary of the $0-$handle, i.e. to the
3-sphere. Therefore we must exchange $\partial A_{cork}K$ by $S^{3}$
to obtain for the $n$th level\begin{equation}
S_{EH}(W_{n})=-L_{P}\sqrt[3]{vol(S^{3}\setminus N(Wh))}\cdot CS(S^{3}\setminus N(Wh))\label{eq:action-Wn}\end{equation}
using the vanishing of $CS(S^{3})=0$. Up to now we do not discuss
the size of $W_{n}$ relative to $W_{n-1}$. As shown by Freedman
\cite{Fre:82}, there is a complex network of re-embedding theorems
so the $7-$stage tower embed into a $6-$stage tower etc. Therefore
we define the relation\begin{equation}
S_{EH}(W_{n})=\sigma\cdot S_{EH}(W_{n-1})\label{eq:action-relation}\end{equation}
with the free (regularization) parameter $0<\sigma<1$ to reflect
the inclusion of higher stage towers into lower stage towers. Finally
we obtain for the action\begin{eqnarray*}
S_{EH}(\mathbb{R}_{\theta}^{4}) & = & \lambda_{A_{cork}}\cdot vol(A_{cork})+L_{P}\sqrt[3]{vol(\partial A_{cork})}\cdot CS(\partial A_{cork})\\
 & - & L_{P}\sqrt[3]{vol(\partial A_{cork}\setminus N(Wh))}\cdot CS(\partial A_{cork}\setminus N(Wh))\\
 & - & L_{P}\sqrt[3]{vol(S^{3}\setminus N(Wh))}\cdot CS(S^{3}\setminus N(Wh))\cdot\frac{\sigma}{1-\sigma}\end{eqnarray*}
where we used the equations (\ref{eq:action-exotic-R4-1},\ref{eq:action-K},\ref{eq:action-W1},\ref{eq:action-Wn},\ref{eq:action-relation})
and the series\[
\frac{\sigma}{1-\sigma}=\sum_{i=1}^{\infty}\sigma^{i}\]
It is obvious that the complexity of the Casson handle is encoded
in the series of $\sigma$ determined by the relation (\ref{eq:action-relation}).
Here we assume a Casson handle with no branching. In the general case
we have to include the branching at every level. Then one can encode
the tree structure into a general polynomial\[
\sum_{n=1}^{\infty}a_{n}\sigma^{n}\]
where the coefficients $a_{n}$ encode the branching information,
i.e. $a_{n}$ is the number of branchings at the level $n$. Then
we have to choose $\sigma$ so that the sum converges. We will later
come back to this point. 

Before we will discuss the path integral, some words about the values
of the Chern-Simons invariants.. At first the Akbulut cork $K$ has
the boundary $\partial K=\Sigma(2,5,7)$, a Brieskorn sphere. The
(minimal) Chern-Simons invariant of this homology sphere was calculated
in\cite{FinSte:90,KirKla:90,FreGom:91} to be\[
CS(\Sigma(2,5,7))=\frac{9}{280}\bmod1\]
Usually there is more than one value but only the minimal one corresponds
to the Levi-Civita connection (needed in our Einstein-Hilbert action).
Using SnapPea of J. Weeks one can also calculate the volume and the
Chern-Simons invariant of the complement $S^{3}\setminus N(Wh)$.
This complement is a hyperbolic 3-manifold. By Mostow rigidity \cite{Mos:68}
the volume and the Chern-Simons invariant are topological invariants.
Then one obtains for the complement of the Whitehead link \begin{eqnarray*}
vol(S^{3}\setminus N(Wh)) & = & 3.66386...\\
CS(S^{3}\setminus N(Wh)) & = & +2.46742...\bmod\pi^{2}\end{eqnarray*}
For the complement $\partial A_{cork}\setminus N(Wh)$ (obviously
it is a hyperbolic 3-manifold) we use a sum decomposition\[
\partial A_{cork}\setminus N(Wh)=\partial A_{cork}\#S^{3}\setminus N(Wh)\]
and get for the Chern-Simons invariant\begin{eqnarray*}
CS(\partial A_{cork}\#S^{3}\setminus N(Wh)) & = & CS(\partial A_{cork})+CS(S^{3}\setminus N(Wh))\\
 & = & CS(\Sigma(2,5,7))+CS(S^{3}\setminus N(Wh))\end{eqnarray*}
Mostow rigidity means also that we have to introduce a fixed length
scale $L_{W}$ for $W_{1},W_{2}$ which will be scaled for $W_{n}$
by $\sigma$. Then in the usual units we obtain the action\begin{eqnarray}
\frac{1}{\hbar}S_{EH}(\mathbb{R}_{\theta}^{4}) & = & \frac{1}{\hbar}S_{EH}(\mathbb{R}^{4})+\frac{\lambda_{A_{cork}}}{L_{P}^{2}}\cdot vol(A_{cork})+\frac{\sqrt[3]{vol(\partial A_{cork})}}{L_{P}}\cdot CS(\partial A_{cork})\nonumber \\
 & - & \frac{L_{W}\cdot\sqrt[3]{vol(\partial A_{cork}\setminus N(Wh))}}{L_{P}^{2}}\cdot\left(CS(\partial A_{cork})+CS(S^{3}\setminus N(Wh))\right)\nonumber \\
 & - & \frac{L_{W}\cdot\sqrt[3]{vol(S^{3}\setminus N(Wh))}}{L_{P}^{2}}\cdot CS(S^{3}\setminus N(Wh))\cdot\frac{\sigma}{1-\sigma}\label{eq:action-small-exotic-R4}\end{eqnarray}
for the small exotic $\mathbb{R}_{\theta}^{4}$.

\section{The functional integral\label{sec:The-functional-integral}}

Now we will discuss the (formal) path integral\begin{equation}
Z=\int Dg\ \exp\left(\frac{i}{\hbar}S_{EH}[g]\right)\label{eq:path-integral-metric}\end{equation}
with the action (\ref{eq:EH-action}) and its conjectured dependence
on the choice of the smoothness structure. In the following we will
using frames $e$ or connections $\Gamma$ instead of the metric $g$.
Furthermore we will ignore all problems (ill-definiteness, singularities
etc.) of the path integral approach. Then instead of (\ref{eq:path-integral-metric})
we have\[
Z=\int De\:\exp\left(\frac{i}{\hbar}S_{EH}[e,M]\right)\]
with the action\[
S_{EH}[e,M]=\intop_{M}tr(e\wedge e\wedge R)\]
where $e$ is a 1-form (coframe), $R$ is the curvature 2-form $R$
and $M$ is the 4-manifold. Next we have to discuss the measure $De$
of the path integral. Currently there is no rigorous definition of
this measure and as usual we assume a product measure. The calculation
(\ref{eq:relation-actions-large-exotic-R4},\ref{eq:action-small-exotic-R4})
of the action for the two types of exotic $\mathbb{R}^{4}$ (large
and small) shows an expected independence of the exotic smoothness
from a metric. Exotic smoothness of the large/small $\mathbb{R}^{4}$
depends only on a continuous parameter (say $t$) and the path integral
(as ''sum over geometries and differential structure'')
should integrate over this parameter as well. 

So the action $S_{EH}(R)$ of an exotic $\mathbb{R}^{4}$ (denoted
by $R^{4}$) has the principal structure\[
S_{EH}[e,R^{4}]=S_{EH}[e,\mathbb{R}^{4}]+S_{exotic}[t]\]
and we obtain\begin{eqnarray*}
Z & = & \int De\:\exp\left(\frac{i}{\hbar}S_{EH}[e,R^{4}]\right)\\
 & = & \left(\int De\,\exp\left(\frac{i}{\hbar}S_{EH}[e,\mathbb{R}^{4}]\right)\right)\cdot\int Dt\:\exp\left(\frac{i}{\hbar}S_{exotic}[t]\right)\end{eqnarray*}
for the path integral. The first part\begin{equation}
Z_{0}=\intop_{Geometries}De\:\exp\left(\frac{i}{\hbar}S_{EH}[e,\mathbb{R}^{4}]\right)\label{eq:geometrical-part-state-sum}\end{equation}
is the formal integration over the geometries and we are left with
the second integration\begin{equation}
\int Dt\:\exp\left(\frac{i}{\hbar}S_{exotic}[t]\right)\label{eq:exotic-part-state-sum}\end{equation}
by varying the differential structure of $\mathbb{R}^{4}$. As explained
above, there are two possible classes of exotic $\mathbb{R}^{4}$,
the large and the small exotic $\mathbb{R}^{4}$. So the parameter
has a different meaning. In the large case, the parameter is the radius
of the continuous family of large exotic $\mathbb{R}^{4}$ eliminating
the geometrical dependence in the action (\ref{eq:relation-actions-large-exotic-R4})
(see the discussion below). Because of the Mostow rigidity \cite{Mos:68},
the contribution of the small exotic $\mathbb{R}^{4}$ to the action
(\ref{eq:action-small-exotic-R4}) is purely topological (except for
the volume $vol(K)$ giving only a numerical shift in the action).
Finally we have,

\begin{proposition}

The (formal) path integral $Z$ of an exotic $\mathbb{R}^{4}$ splits
into a product of two path integrals\begin{equation}
Z=\int De\:\exp\left(\frac{i}{\hbar}S_{EH}[e,R^{4}]\right)=Z_{0}\cdot\int Dt\:\exp\left(\frac{i}{\hbar}S_{exotic}[t]\right)\label{eq:splitting-state-sum-exotic-R4}\end{equation}
i.e. the exotic part is independent of a frame or metric.

\end{proposition}

\subsection{Large exotic $\mathbb{R}^{4}$}

In subsection \ref{sub:Large-exotic-R4} we discuss the construction
of a large exotic $\mathbb{R}^{4}$ by using a \textbf{fixed} topologically
but non-smoothly sliced knot $K$. Thus by using the relation (\ref{eq:relation-actions-large-exotic-R4})
we obtain\begin{equation}
Z=Z_{0}\cdot\exp\left(-i\frac{vol(D_{2})}{L_{P}^{2}}\cdot4\pi^{2}\cdot g_{4}(K)\right)\label{eq:state-sum-large-exotic-R4}\end{equation}
Here we obtain only countable many large exotic $\mathbb{R}^{4}$
in this way. To distinguish uncountable many exotic $\mathbb{R}^{4}$'s
one has to use the following construction. Let $R_{K}^{4}$ be an
exotic $\mathbb{R}^{4}$ constructed from a topologically but non-smoothly
sliced knot $K$. Now fix a homeomorphism $h:\mathbb{R}^{4}\to R_{K}^{4}$
and let $R_{r}\subset R_{K}^{4}$ be the image of the open balls of
radius $r$ centered at $0$ in $\mathbb{R}^{4}$ with $R_{\infty}=R_{K}^{4}$.
Each $R_{t}$ inherits a smooth structure as an open subset of $R_{K}^{4}$
\cite{Qui:82}. By the work of Freedman and Quinn \cite{FreQui:90}
any homeomorphism between smooth 4-manifolds is isotopic to one which
is a local diffeomorphism near a preassigned 1-complex (one axis e.g.
the non-negative $x_{1}-$axis). Using this result, one can show (see
Theorem 9.4.10 in \cite{GomSti:97}) that $R_{s}$ and $R_{t}$ are
non-diffeomorphic for $0<s<t<\infty$. The proof based on the fact
that there is a compact 4-manifold $K\subset R_{t}$ which cannot
be embedded in $R_{s}$. In the construction of subsection \ref{sub:Large-exotic-R4}
we used a decomposition (\ref{eq:decomposition-large-exotic-R4})
with respect to a topologically flat embedding $\rho:X_{K}\to\mathbb{R}^{4}$
of a 2-handle body $X_{K}$. This submanifold $X_{K}$ is compact
and the exotic smoothness structure is determined by the smooth failure
of this embedding. But then the 4-ball $D^{4}\subset X_{K}$ cannot
be embedded. Conversely the image $\rho(X_{K})$ is a compact manifold
as well and can be surrounded by a 4-ball $\rho(X_{K})\subset D^{4}$
(Heine-Borel theorem). Therefore it is enough to consider a scaling
of $X_{K}$ by the radius $t$ to express the radius family $R_{t}$
defined above. In the state sum (\ref{eq:state-sum-large-exotic-R4})
above, one has to replace $vol(D_{2})$ by $t^{2}$ to express this
scaling. For a fixed knot $K$ (topologically slice but smoothly non-slice)
and a fixed parameter we have the state sum\[
Z_{t}=Z_{0}\cdot\exp\left(-i\frac{t^{2}}{L_{P}^{2}}\cdot4\pi^{2}\cdot g_{4}(K)\right)\]
For the full path integral we obtain\[
Z_{large}=Z_{0}\cdot\frac{1}{L_{P}}\intop_{0}^{\infty}\exp\left(-i\frac{t^{2}}{L_{P}^{2}}\cdot4\pi^{2}\cdot g_{4}(K)\right)dt\]
in units of the Planck length and finally (using the Fresnel integral)\begin{equation}
Z_{large}=Z_{0}\cdot\frac{e^{-i\pi/4}}{2\pi\sqrt{g_{4}(K)}}\label{eq:state-sum-all-large-exotic-R4}\end{equation}
Especially the contribution of the large exotic $\mathbb{R}^{4}$
is independent of a scale, i.e. it is (differential-)topological invariant.

\subsection{Small exotic $\mathbb{R}^{4}$}

Now we use our technique to represent the small exotic $\mathbb{R}^{4}$
using the decomposition of subsection \ref{sub:Small-exotic-R4}.
The final result for the action was formula (\ref{eq:action-small-exotic-R4}).
Then we can read the expression for the state sum\begin{eqnarray}
Z & = & Z_{0}\cdot\exp\left(i\frac{\lambda_{A_{cork}}}{L_{P}^{2}}\cdot vol(A_{cork})+i\cdot\Lambda_{\partial A_{cork}}\cdot CS(\partial A_{cork})\right)\cdot\nonumber \\
 &  & \cdot\exp\left(-i\cdot\Lambda_{\partial A_{cork},Wh}\cdot\left(CS(\partial A_{cork})+CS(S^{3}\setminus N(Wh))\right)\right)\nonumber \\
 &  & \cdot\exp\left(-i\cdot\Lambda_{S^{3},Wh}\cdot CS(S^{3}\setminus N(Wh))\cdot\frac{\sigma}{1-\sigma}\right)\label{eq:state-sum-small-exotic-R4-simplest-CH}\end{eqnarray}
with the scaling parameters\begin{eqnarray}
\Lambda_{\partial A_{cork}} & = & \frac{\sqrt[3]{vol(\partial A_{cork})}}{L_{P}}\nonumber \\
\Lambda_{\partial A_{cork},Wh} & = & \frac{L_{W}\cdot\sqrt[3]{vol(\partial A_{cork}\setminus N(Wh))}}{L_{P}^{2}}\label{eq:scaling-parameters}\\
\Lambda_{S^{3},Wh} & = & \frac{L_{W}\cdot\sqrt[3]{vol(S^{3}\setminus N(Wh))}}{L_{P}^{2}}\nonumber \end{eqnarray}
As remarked above, the Whitehead link $Wh$ is an hyperbolic link,
i.e. the knot complements $S^{3}\setminus N(Wh)$ and $\partial A_{cork}\setminus N(Wh)$
are hyperbolic 3-manifolds (with boundary the disjoint union of two
tori). It is not an unexpected result that the contribution of the
small exotic $\mathbb{R}^{4}$ is a topological invariant (Chern-Simons
invariant) by fixed scaling parameters $\Lambda$. The variation of
the Casson handle produces other small exotic $\mathbb{R}^{4}$. The
whole problem was analyzed in \cite{DeMichFreedman1992} using the
so-called design of a Casson handle, i.e. a singular parametrization
of all Casson handles by a binary tree. As shown by Freedman \cite{Fre:82},
the design forms a continuous set (Cantor continuum)%
\footnote{This kind of Cantor set is given by the following construction: Start
with the unit Interval $S_{0}=[0,1]$ and remove from that set the
middle third and set $S_{1}=S_{0}\setminus(1/3,2/3)$ Continue in
this fashion, where $S_{n+1}=S_{n}\setminus\left\{ \mbox{middle thirds of subintervals of \ensuremath{S_{n}}}\right\} $.
Then the Cantor set $C.s.$ is defined as $C.s.=\cap_{n}S_{n}$. With
other words, if we using a ternary system (a number system with base
3), then we can write the Cantor set as all sequences containing only
$0$ or $2$ after the decimal point.%
}. Then to every real number%
\footnote{According to \cite{DeMichFreedman1992}, there is a collection of
parameter values (representing the Casson handle) with the cardinality
of the continuum in the Zermelo-Fraenkel set theory with choice so
that the corresponding small exotic $\mathbb{R}^{4}$'s are pairwise
non-diffeomorphic.%
} in $[0,1]$ there is a Casson handle. In the formula above, the simplest
Casson handle is represented by the expression $\frac{\sigma}{1-\sigma}$
and we have to replace it by the real number $t\in[0,1]$. Then the
last term in (\ref{eq:state-sum-small-exotic-R4-simplest-CH}) is
now replaced by\begin{eqnarray*}
\intop_{0}^{1}\exp\left(-i\cdot\Lambda_{S^{3},Wh}\cdot CS(S^{3}\setminus N(Wh))\cdot t\right)dt & =\\
\frac{e^{i\pi/2}\left(\exp\left(-i\Lambda_{S^{3},Wh}\cdot CS(S^{3}\setminus N(Wh))\right)-1\right)}{\Lambda_{S^{3},Wh}\cdot CS(S^{3}\setminus N(Wh))}\end{eqnarray*}
and therefore we obtain finally\begin{eqnarray}
Z_{small} & = & Z_{0}\cdot\exp\left(i\frac{\lambda_{A_{cork}}}{L_{P}^{2}}\cdot vol(A_{cork})+i\cdot\Lambda_{\partial KA_{cork}}\cdot CS(\partial A_{cork})\right)\cdot\nonumber \\
 &  & \cdot\exp\left(-i\cdot\Lambda_{\partial A_{cork},Wh}\cdot\left(CS(\partial A_{cork})+CS(S^{3}\setminus N(Wh))\right)\right)\nonumber \\
 &  & \cdot\frac{e^{i\pi/2}\left(\exp\left(-i\Lambda_{S^{3},Wh}\cdot CS(S^{3}\setminus N(Wh))\right)-1\right)}{\Lambda_{S^{3},Wh}\cdot CS(S^{3}\setminus N(Wh))}\label{eq:state-sum-small-exotic-R4}\end{eqnarray}
Then the contribution of the small exotic $\mathbb{R}^{4}$ is independent
of a scale again, i.e. it is (differential-)topological invariant. 

\begin{proposition}

Exotic smoothness contributes to the state sum of quantum gravity
for all exotic $\mathbb{R}^{4}$.

\end{proposition}

\section{Observables\label{sec:Observables}}

Any consideration of quantum gravity is incomplete without considering
observables and its expectation values. Here we consider two kinds
of observables:
\begin{enumerate}
\item Volume 
\item holonomy along open and closed paths (Wilson loop)
\end{enumerate}
The expectation value for the volume can be calculated only for compact
submanifolds of the $\mathbb{R}^{4}$ by using the decomposition of
the large $R_{K}^{4}$ or small $\mathbb{R}_{\theta}^{4}$ exotic
$\mathbb{R}^{4}$\begin{eqnarray*}
R_{K}^{4} & = & \left(\mathbb{R}^{4}\setminus int\rho(X_{K})\right)\cup_{\partial X_{K}}X_{K}\\
\mathbb{R}_{\theta}^{4} & = & int\left(A_{cork}\cup_{N}W\cup_{N}W\cup_{N}\cdots\right)\end{eqnarray*}
explained in section \ref{sec:Exotic-R4}. Let $D\subset M$ be a
submanifold in $M=R_{K}^{4},\mathbb{R}_{\theta}^{4}$ with volume
$vol(D)$ (where its meaning depends on the dimension of $D$). Let
\[
\left\langle Vol(D)\right\rangle _{0}=\frac{\int De_{G}\: Vol(D,e_{G})\exp\left(\frac{i}{\hbar}S_{EH}[e,M]\right)}{\int De_{G}\:\exp\left(\frac{i}{\hbar}S_{EH}[e,M]\right)}\]
be the expectation value of the volume w.r.t. the geometry. At first
we assume that $D$ is 1-dimensional, i.e. $D=[0,1]$ or $D=S^{1}$.
In this case, there is no topological restriction and we obtain any
value of $vol(D)$ (remember $M$ is simple connected and so every
loop is contractable). The case of a surface is the first non-trivial
example. So, we consider a closed surface $D$ of genus $g$ (the
surface with boundary can be simply obtained from this case by removing
disks). For the large exotic $\mathbb{R}^{4}$ denoted by $R_{K}^{4}$
we have to consider two cases: 
\begin{enumerate}
\item $D$ lies at $X_{K}$ or
\item $D$ lies at $\mathbb{R}^{4}\setminus int\rho(X_{K})$
\end{enumerate}
In the first case, the surface $D$ is confined by the handle body
$X_{K}$. By using the action (\ref{eq:relation-actions-large-exotic-R4})
we can express the expectation value of the volume by a formal variation
of $1/L_{P}^{2}$ to get\begin{eqnarray*}
\left\langle Vol(D)\right\rangle _{0} & = & \frac{\delta\ln Z}{\delta(1/L_{P}^{2})}=\frac{1}{Z}\int De_{G}\:\frac{i\partial S_{EH}}{\partial(1/L_{P}^{2})}\exp\left(\frac{i}{\hbar}S_{EH}[e,M]\right)\\
 & = & vol(D_{2})4\pi^{2}g_{4}(K)\end{eqnarray*}
The disk is confined by $X_{K}$ as expected. Especially the whole
effect vanishes for the standard $\mathbb{R}^{4}$ with $g_{4}(K)=0$.
Furthermore for the radial family of large exotic $\mathbb{R}^{4}$,
the volume $vol(D_{2})$ is related to the radius parameter $r$ by
$r^{2}=vol(D_{2})$. Then the size of $vol(D_{2})$ is fixed for one
smoothness structure (up to diffeomorphisms), i.e. \emph{we obtain
a quantized area of a surface} $D$ in units of $vol(D_{2})\cdot g_{4}(K)\cdot4\pi^{2}$.
For the second case we do not get any restriction on the volume (or
better area) of $D$.

There is a similar effect for the small exotic $\mathbb{R}^{4}$.
Here the situation is more complicate. For instance, the surface $D$
can be inside of the $\partial A_{cork}$ or in one of the periodic
pieces $W_{i}$. But the three scaling parameters $\Lambda$ defined
by (\ref{eq:scaling-parameters}) must be quantized. Here one argues
via a consistent quantum field theory based on the Chern-Simons action
(see Witten \cite{Wit:91.2}) to show the quantization of the parameters.
As an example we consider the quantized parameter\[
\Lambda_{S^{3},Wh}=\frac{L_{W}\cdot\sqrt[3]{vol(S^{3}\setminus N(Wh))}}{L_{P}^{2}}\in\mathbb{N}\]
The volume $vol(S^{3}\setminus N(Wh))$ is a topological invariant
by Mostow rigidity (the Whitehead links is hyperbolic). Therefore
the length scale $L_{W}$ of the periodic pieces are also quantized.
But we remark that for all 1-dimensional submanifolds there is no
restriction, i.e. the length of these submanifolds is not quantized.
Finally we obtain:

\begin{proposition}\label{prop:area-volume-quanta}The area and the
volume of a 3-dimensional submanifold must be quantized in units of
$\Lambda\cdot L_{P}^{2}$ or $\Lambda\cdot L_{P}^{3}$, respectively.
The factor $\Lambda$ depends on the concrete smoothness structure.
\end{proposition}

Now we discuss the holonomy \[
hol(\gamma,\Gamma)=Tr\left(\exp\left(i\intop_{\gamma}\Gamma\right)\right)\]
along a path $\gamma$ w.r.t. the connection $\Gamma$, as another
possible observable. The spacetime $M$ is simple-connected. Thus
every path can be deformed to another path. Especially, every closed
path is the boundary of a disk (every knot in a 4-space is smoothly
deformable (=isotopic) to the unknot). The embedding of this disk
is the non-trivial task, i.e. an immersed disk is the best possible
alternative to an embedding. But then this disk has self-intersections
in the interior which are the source of singularities. Another possibility
is the existence of a surface $F$ (the Seifert surface) of non-zero
genus with $\partial F$ equal to the closed curve. Usually the exotic
$\mathbb{R}^{4}$ splits into 4-dimensional pieces $W$ with $\partial W\not=\emptyset$.
Therefore we embed (or immerse) the \emph{closed curve} $\gamma=\partial F$
into $\partial W$ and the surface $F$ into $W$. For the calculation
of the expectation value, we have to integrate over the connections
in the path integral\[
\left\langle W(\gamma)\right\rangle =\frac{\int D\Gamma\: W(\gamma)\exp\left(\frac{i}{\hbar}\, S_{EH}(R^{4},\Gamma)\right)}{\int D\Gamma\:\exp\left(\frac{i}{\hbar}\, S_{EH}(R^{4},\Gamma)\right)}\]
with the Wilson loop\[
W(\gamma)=Tr\left(\exp\left(i\intop_{\gamma}\Gamma\right)\right)\]
for the exotic $R^{4}$. For large exotic $\mathbb{R}^{4}$'s we do
not find any interesting value for the expectation value $\left\langle W(\gamma)\right\rangle $
except the usual expression for standard $\mathbb{R}^{4}$. The reason
could be that the action of the large exotic $\mathbb{R}^{4}$ is
only corrected by a constant term in comparison to the action of the
standard $\mathbb{R}^{4}$. From the classical point of view, only
solutions with a non-trivial curvature will be preferred or the large
exotic $\mathbb{R}^{4}$ is always curved (in agreement with \cite{Sladkowski2001}).
But this effect cannot be localized at some holonomy along closed
curve. In contrast small exotic $\mathbb{R}^{4}$ have different properties.
Now we have two different contributions:
\begin{enumerate}
\item The closed curve lies at $\partial A_{cork}$ (and the Seifert surface
in $A_{cork}$), or
\item the closed curve lies at $\partial W$ (and the Seifert surface in
$W$).
\end{enumerate}
The first case leads to an integral\[
\left\langle W(\gamma)\right\rangle =\frac{\int D\Gamma\: W(\gamma)\exp\left(i\cdot(\Lambda_{\partial A_{cork}}-\Lambda_{\partial A_{cork},Wh})\cdot CS(\partial A_{cork},\Gamma)\right)}{\int D\Gamma\:\exp\left(\frac{i}{\hbar}\, S_{EH}(R^{4},\Gamma)\right)}\]
so that we obtain a knot invariant \[
\left\langle W(\gamma)\right\rangle =\mbox{generalized Jones polynomial }J_{\gamma}(q)\]
of the closed curve $\gamma$ in $\partial A_{cork}$ where we have
\[
q=\exp\left(\frac{2\pi i}{2+(\Lambda_{\partial A_{cork}}-\Lambda_{\partial A_{cork},Wh})}\right)\]
The second case is very similar and we obtain the principal result\[
\left\langle W(\gamma)\right\rangle =\frac{\int D\Gamma\: W(\gamma)\exp\left(i\cdot\ell\cdot CS(S^{3}\setminus N(Wh),\Gamma)\right)}{\int D\Gamma\:\exp\left(\frac{i}{\hbar}\, S_{EH}(R^{4},\Gamma)\right)}\]
where the constant $\ell$ is a combination of the scaling parameters.
But the corresponding quantum field theory is only consistent, if
the coefficients of the Chern-Simons terms are integer valued. This
fact confirms again the Proposition \ref{prop:area-volume-quanta}.

\section{Naked singularities and the failure of the Whitney trick\label{sec:Naked-singularities}}

In this section we will discuss the appearance of naked singularities
in exotic $\mathbb{R}^{4}$. The Cauchy surface of the standard $\mathbb{R}^{4}$
is given by $\mathbb{R}^{3}$ so that $\mathbb{R}^{4}=\mathbb{R}^{3}\times\mathbb{R}^{1}$
in agreement with the discussion above. In contrast, any exotic $\mathbb{R}^{4}$
cannot be split like (3-manifold$\times\mathbb{R}$) . 

\begin{figure}
\includegraphics[angle=90,scale=0.5]{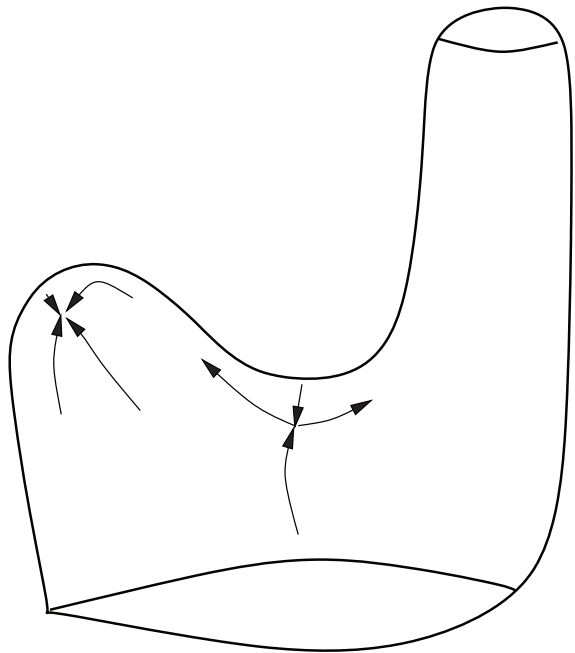}

\caption{two naked singularities\label{fig:toy-naked-singularity}}

\end{figure}

\begin{figure}
\includegraphics{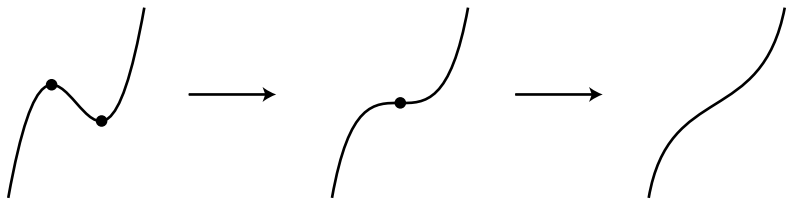}

\caption{killling a 0- and a 1-handle\label{fig:killling-handles}}

\end{figure}

To visualize the problem, we consider the following toy model: a non-trivial
surface (see Fig. \ref{fig:toy-naked-singularity}) connecting two
circles which can be deformed to the usual cylinder. %
 This example can be described by the concept of a cobordism. A cobordism
$(W,M_{1},M_{2})$ between two $n-$manifolds $M_{1},M_{2}$ is a
$(n+1)-$manifold $W$ with $\partial W=M_{1}\sqcup M_{2}$ (ignoring
the orientation). Then there exists a smooth function $f:W\to[0,1]$
such that $f^{-1}(0)=M_{1},\, f^{-1}(1)=M_{2}$. By general position,
one can assume that $f$ is a Morse function and such that all critical
points occur in the interior of $W$. In this setting $f$ is called
a Morse function on a cobordism. For every critical point of $f$
(vanishing first derivative) one adds a handle $D^{k}\times D^{n-k}$.
In our example in Fig. \ref{fig:toy-naked-singularity}, we add a
2-handle $D^{2}\times D^{0}$ (the maximum) and a 1-handle $D^{1}\times D^{1}$
(the saddle). But obviously this cobordism is diffeomorphic to the
trivial one $S^{1}\times[0,1]$ because the two boundary components
are diffeomorphic to each other. Therefore the 2-/1-handle pair is
''killed'' in this case. The 2-handle
and the 1-handle differ in one direction where the Morse function
has a maximum for the 2-handle and a minimum for the 1-handle. The
left graph of Fig. \ref{fig:killling-handles} visualizes this fact.
 Furthermore the sequence of graphs from the left to right presents
the process to ''kill'' the handle
pair. In the definition of the cobordism, there is no restriction
on the two manifolds $M_{1},M_{2}$, i.e. one can consider a cobordism
between two non-homeomorphic manifolds. An example is a cobordism
between one circle $S^{1}$ and the disjoint union of two circles
$S^{1}\sqcup S^{1}$ (the pair of pants). But in the discussion above
(see section \ref{sec:Lorentz-metric-global-hyp}) we considered always
a special class of cobordisms, where the two manifolds $M_{1},M_{2}$
are homeomorphic to each other. Mathematically we have to discuss
h-cobordisms between 3-manifolds. Because of the homeomorphism between
$M_{1}$ and $M_{2}$, the h-cobordism must be contain any handle
in the interior. Usually the construction of a h-cobordism will produce
also handles in the interior. But these handles can be killed where
the details of the construction can be found in \cite{Mil:65}. Here
we will give only some general remarks. Any $0-/1-$handle pair as
well any $n-/(n+1)-$handle pair (remember the h-cobordism is $n+1$-dimensional)
can be killed by a general procedure. The killing of a $k-/(k+1)-$handle
pair depends on a special procedure, the Whitney trick. For 4- and
5-dimensional h-cobordisms (between 3- and 4-manifolds, respectively)
we cannot use the Whitney trick. This failure lies at the heart of
the problem to classify 3- and 4-manifolds.

Now we will specialize to a 4-dimensional h-cobordism between 3-manifolds.
Then we can kill the $0-/1-$and the $3-/4-$handle pair of the h-cobordism.
Then we are left with pairs of $2-$handles. If the Whitney trick
works in this case, we can kill these pairs of handles. But it is
known that the Whitney trick only works topologically. But the existence
of exotic $S^{3}\times\mathbb{R}$'s (as non-compact examples) gave
counterexamples, so that the pairs of 2-handles never cancel each
other. The critical point of the Morse function with index $2$ (the
Morse function has a minimum in two directions (saddle point)) corresponds
to the $2-$handle. Each pair of 2-handles is connected to each other,
i.e. the directions representing the minimum of a 2-handle are connected
with the directions representing the maximum of the other 2-handle.
Therefore we get

\begin{proposition}

The naked singularities of an exotic $\mathbb{R}^{4}$ are pairs of
2-handles which cancel topologically but not smoothly by the failure
of the Whitney trick.

\end{proposition}

\section{Cosmological consequences}

The global structure of the spacetime $\mathbb{R}^{4}$ is greatly
influenced by the smoothness structure. Therefore it seems natural
to obtain cosmological results from our calculation of the functional
integral above. For the large exotic $\mathbb{R}^{4}$ we obtained
the term (see (\ref{eq:relation-actions-large-exotic-R4}))\[
-\frac{vol(D_{2})}{L_{P}^{2}}\cdot4\pi^{2}\cdot g_{4}(K)\]
as the correction of the action $S_{EH}(\mathbb{R}^{4})$. This term
can be simply interpreted as the cosmological constant term. At first
we remark that the spacetime $\mathbb{R}^{4}$ contains the big bang
singularity. The removement of this singularity (say at $0$) leads
to\[
\mathbb{R}^{4}\setminus\left\{ 0\right\} =S^{3}\times\mathbb{R}\]
i.e. we assume a compact spatial component $S^{3}$ of the cosmos.
But then we have set for the term above\[
-\frac{vol(D_{2})}{L_{P}^{2}}\cdot4\pi^{2}\cdot g_{4}(K)=\intop_{S^{3}\times[0,1]}\Lambda_{cosmo}\sqrt{g}d^{4}x\]
where we integrate over the time period since the big bang, i.e. the
cosmological constant is given by \[
\Lambda_{cosmos}=-\frac{vol(D_{2})}{L_{P}^{2}\cdot vol(S^{3}\times[0,1])}\cdot4\pi^{2}\cdot g_{4}(K)<0\]
a negative number.

The case of a small exotic $\mathbb{R}^{4}$ is more interesting.
The action (\ref{eq:action-small-exotic-R4}) contains the term \[
\frac{\lambda_{A_{cork}}}{L_{P}^{2}}\cdot vol(A_{cork})\]
which can be similarly interpreted as cosmological constant term.
In contrast to the large exotic $\mathbb{R}^{4}$, we have now a direct
model of cosmos. The compact submanifold $A_{cork}$ in the construction
of the small exotic $\mathbb{R}^{4}$ can serve as a model for the
cosmic evolution. This submanifold $A_{cork}$ is contractable with
a homology 3-sphere as boundary. Then the cosmological constant is
given by\[
\Lambda_{cosmos}=\lambda_{A_{cork}}\sim\frac{1}{\sqrt{vol(A_{cork})}}\]
Therefore exotic smoothness can be the appropriate view to understand
the dark energy.

\section{Conclusion}

In this paper we discussed the influence of exotic smoothness on the
functional integral of the Einstein-Hilbert action. Then we obtain
a bunch of results:
\begin{itemize}
\item the appearance of naked singularities in exotic $\mathbb{R}^{4}$,
\item any naked singularity is a saddle point (of index 2, i.e. two directions
are a minimum) and we have only an even number of it,
\item area and volume quantization by using Mostow rigidity \cite{Mos:68}
agreeing with results in Loop quantum gravity \cite{RovSmo:95},
\item the appearance of a cosmological constant.
\end{itemize}
This is only the beginning of a systematic analysis of exotic $\mathbb{R}^{4}$'s.
Interestingly, there are also rich connections between quantization,
non-commutative geometry and exotic smoothness \cite{AsselmeyerKrol2010}.
\\
\emph{Finally we can support the physically motivated conjecture
that quantum gravity depends on exotic smoothness.}

\ack{}{This work was partly supported (T.A.) by the LASPACE grant. Many
thanks to Carl H. Brans for nearly infinite many discussions about
the physics of exotic 4-manifolds. The authors acknowledged for all
mathematical discussions with Duane Randall and Terry Lawson. The
section about global hyperbolicity is meanly inspired by the discussion
with Miguel S\'anchez. Many thanks for the enlightened remarks. }

\appendix

\section{Connected and boundary-connected sum of manifolds\label{sec:Connected-and-boundary-connected}}

Now we will define the connected sum $\#$ and the boundary connected
sum $\natural$ of manifolds. Let $M,N$ be two $n$-manifolds with
boundaries $\partial M,\partial N$. The \emph{connected sum} $M\#N$
is the procedure of cutting out a disk $D^{n}$ from the interior
$int(M)\setminus D^{n}$ and $int(N)\setminus D^{n}$ with the boundaries
$S^{n-1}\sqcup\partial M$ and $S^{n-1}\sqcup\partial N$, respectively,
and gluing them together along the common boundary component $S^{n-1}$.
The boundary $\partial(M\#N)=\partial M\sqcup\partial N$ is the disjoint
sum of the boundaries $\partial M,\partial N$. The \emph{boundary
connected sum} $M\natural N$ is the procedure of cutting out a disk
$D^{n-1}$ from the boundary $\partial M\setminus D^{n-1}$ and $\partial N\setminus D^{n-1}$
and gluing them together along $S^{n-2}$ of the boundary. Then the
boundary of this sum $M\natural N$ is the connected sum $\partial(M\natural N)=\partial M\#\partial N$
of the boundaries $\partial M,\partial N$.

\section*{References}

%\bibliographystyle{elsarticle-num-names}
%\bibliography{DIFFBIB,foliation-gerbes}

\end{document}